# Effects of post-anneal conditions on the dielectric properties of $CaCu_3Ti_4O_{12}$ thin films prepared on Pt/Ti/SiO$_2$/Si substrates

Liang Fang [a] and Mingrong Shen

Department of Physics, Suzhou University, Suzhou 215006, People's Republic of China

Wenwu Cao [b]

Department of Physics and Materials Science, City University of Hong Kong, Kowloon, Hong Kong, China

**ABSTRACT:**

High-dielectric-constant $CaCu_3Ti_4O_{12}$ (CCTO) thin films were prepared on Pt/Ti/SiO$_2$/Si(100) substrates by pulsed-laser deposition. The 480 nm thick polycrystalline films have preferred orientation and show obvious crystallization on the surface. The temperature-dependence of dielectric constant and loss of the Pt/CCTO/Pt capacitors is comparable with that of epitaxial CCTO films grown on oxides substrates. We found that the dielectric properties are very sensitive to the post-annealing atmosphere and temperature. Post-annealing in nitrogen atmosphere produces larger low-frequency dielectric relaxation as the annealing temperature increases, while annealing in oxygen atmosphere at high temperature suppresses the relaxation but lowers the dielectric constant. Such results are attributed to the presence of insulating grain boundary barrier layers.

**PACS:** 77.22.Gm, 81.15.Fg, 73.40.Rw

______________________________________________

a) Author to whom correspondence should be addressed; Electronic mail: lfang@suda.edu.cn
b) On leave from the Pennsylvania State University, USA.

With the trend of size reduction of many microelectronic devices, high-dielectric-constant oxides have become increasingly important in microelectronics. Recently, much attention has been paid to an unusual cubic perovskite material $CaCu_3Ti_4O_{12}$ (CCTO).[1-11] Both single crystal and ceramic of CCTO have very high-dielectric constant (in the order of $10^4$) at room temperature and remains almost constant in the temperature range from 100K to 600K. Moreover, from neutron powder diffraction[2] and high resolution X-ray diffraction[3] no evidence of any structural phase transition was found in CCTO from 100K to 600K, which is very desirable for practical device applications.[1]

Some groups have successfully grown high-quality epitaxial CCTO films on oxide substrates.[12,13] These films show single crystal structure and the dielectric constants are larger than 1500 in a wide temperature range. This investigation concentrates on CCTO thin films deposited directly onto Si substrates. Such films could be used in practical large-scale integrated circuits since they are compatible with semiconductor technology. In addition, we will study the effects of post-annealing in different atmosphere and at different temperatures on the dielectric properties.

CCTO pellets used as PLD targets were prepared by mixed oxide method. High purity $CaCO_3$, $TiO_2$ and $CuO$ powders were weighted in appropriate ratio, mixed by ball milling for 10-16 hrs, calcined at 750ºC for 2 hrs, and then pressed into pellets. The pellets were finally fired at 1100ºC for 2 hrs. CCTO thin films with 480 nm in thickness were made on $Pt/Ti/SiO_2/Si(100)$ substrates by the PLD technique.[14] A 248 nm KrF excimer laser (Lambda Physik 105i) operated at 5Hz was focused on a high-density ceramic target of CCTO with an energy density of 2 J cm$^{-2}$. The chamber was firstly pumped down to $2\times10^{-6}$ Torr, then oxygen was introduced to a pressure of 200 mTorr. The substrate was heated to 720ºC by a resistance heater during deposition. Using radio-frequency (RF) sputtering technique, Pt electrodes with a diameter of 0.28 mm were deposited onto the top surface of the films at room temperature through a shadow mask. After deposition of the top electrodes, the samples were given different post-anneal treatments in a tube furnace. The temperature dependence of the dielectric properties was measured by a HP4284 LCR meter from 100K to 350K in a Delta 9023 oven. The frequency-dependence of the dielectric properties was measured using a HP4294A impedance analyzer over a frequency range of 100Hz to 2MHz.

Figure 1 shows the XRD patterns of the CCTO target and the CCTO thin film on Pt/Ti/SiO$_2$/Si(100) substrates by a Rigaku D/MAX 3C X-ray Diffraction(XRD) diffractometer using Cu-K$\alpha$ radiation at 40kV. It can be seen that (211), (220), (400) and (422) peaks match the characteristic of the CCTO compounds, demonstrating a polycrystalline characteristic. In addition, comparison of the peak intensities of the film and the target shows that the ratio I(220)/I(422) is higher in the film, indicating a (220) preferential orientation for the CCTO thin film on Pt/Ti/SiO$_2$/Si substrates.

The surface morphology of the CCTO film was analyzed by a Hitachi S-5750 SEM, as shown in the insert of Figure 2(a). The film surface is crack-free and quite smooth. There are many square shaped grains stacking onto each other showing on the film surface, which is consistent with the XRD result indicating (220) preferential orientation. The cross-sectional SEM morphology is presented in the insert of Figure 2(b). The dense CCTO thin films grow column-like, and the interface between the CCTO film and Pt electrode was clean and sharp, which implies negligible inter-diffusion across the interface.

The temperature dependence of the dielectric constant $\varepsilon(T)$ and the loss tangent $tan\delta(T)$ under different frequency for the as-deposited Pt/CCTO/Pt capacitors is shown in figures 2(a) and 2(b), respectively. The dielectric constant has a fairly high value of about 2000 and the loss tangent is less than 0.5 at room temperature below 10kHz and show obvious frequency dispersion. The dielectric constant decreases rapidly with decreasing temperature in the low temperature region (this drop in $\varepsilon(T)$ below 10kHz is not shown here due to the limited temperature range of our measurements). Moreover, there is a broad peak in $tan\delta(T)$ at 1MHz corresponding to the sharp drop of $\varepsilon(T)$, and the temperature peak of the dissipation shifts to lower temperatures with the decrease of frequency. The dielectric constant and loss tangent as well as the characteristics of their temperature dependence at different frequencies are comparable with those of the epitaxial CCTO films deposited on oxides

substrates.[13]

Figure 3(a) and 3(b) shows the variations of the dielectric constant and loss tangent as a function of frequency for the as-deposited Pt/CCTO/Pt capacitors and the same film after annealing in $O_2$ or $N_2$ atmospheres at different temperatures. Note the as-deposited capacitors have gone through six different post-annealing processes as illustrated in Fig.3. Low temperature annealing at 200ºC in nitrogen and oxygen shows no significant changes in the dielectric properties. However, very large dielectric relaxation in low frequency region (below 3kHz) appeared when the sample was post-annealed at 400ºC in $N_2$. This relaxation effect disappeared when a subsequent post-annealing was performed in $O_2$ at the same temperature. Larger dielectric relaxation, both for dielectric constant and loss tangent, was observed when the sample was post-annealed at 550ºC in $N_2$ atmosphere. Again, the relaxation phenomenon disappeared after the sample was exposed to $O_2$ at 550ºC for 8h. It appears that annealing in $N_2$ and $O_2$ atmosphere produce counter effects. One must note that the dielectric constant decreased substantially after this $N_2$ and $O_2$ atmosphere annealing cycle. Using XRD and SEM, we found no structural changes for the samples experienced all the above post-annealing processes.

The mechanism behind the high-dielectric constant of the CCTO ceramics and crystals has been studied by several groups.[1-4,8,10] Recently, the study on impedance spectroscopy[6,11] demonstrated that the CCTO ceramics are composed of semiconducting grains separated by insulating grain boundaries. The origin of the semiconductivity in the grains may arise from a small but countable amount of oxygen-loss during the ceramic fabrication. High-dielectric-constant phenomenon is thus attributed to the grain boundary mechanism, as found in internal barrier layer capacitors.[15] The same mechanism is used to explain the high-dielectric-constant phenomenon in Li and Ti doped NiO.[16] The CCTO films in the present study are polycrystalline and composed of many grains, as illustrated in Figs. 1 and 2. The post-annealing effects on the dielectric properties are very similar to those happened in the ferroelectric polycrystalline film capacitors, such as Pt/BaSrTiO$_3$/Pt capacitors.[17,18] Therefore, we proposed that the present CCTO films on Pt/Ti/SiO$_2$/Si substrate are also composed of semiconducting grains and insulating grain boundary. When the film is post-annealed in the reduced atmosphere, such as in $N_2$, oxygen vacancies are generated at relatively high temperature (above 200 $N_2$ and $O_2$ atmosphere ºC) according to the following reaction:

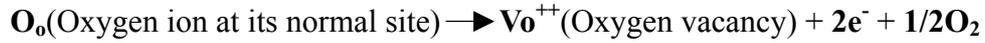
$O_o$(Oxygen ion at its normal site) $\rightarrow$ $Vo^{++}$(Oxygen vacancy) + $2e^-$ + $1/2 O_2$

The oxygen vacancies and space charges (electrons) are produced in the grain boundaries, resulting in low-frequency dielectric relaxation. Higher post-annealing temperature in $N_2$ will produce more oxygen vacancies and space charges, thus, more obvious dielectric relaxation can be observed. On the other hand, when post-annealing in an oxygen atmosphere, the above reaction will be reversed and the oxygen vacancies will be compensated, leading to the disappearance of the dielectric relaxation. Moreover, when the post-annealing temperature is as high as 550ºC, not only oxygen vacancies in the grain boundaries are compensated, but also the oxygen vacancies in part of the originally oxygen-deficient grains near the boundaries. Thus, effectively, the grain boundary layer is increased, causing the dielectric constant to decrease.

In summary, CCTO thin films with high dielectric constant were successfully prepared on Pt/Ti/SiO$_2$/Si (100) substrates by PLD. Microstructure studies reveal that the films are polycrystalline with square shaped crystals on the surface. The dielectric properties and their temperature dependence for the Pt/CCTO/Pt capacitors are comparable with those of epitaxial CCTO films deposited on oxides substrates, and were found to be very sensitive to the post-annealing atmosphere and temperature. Our study demonstrates that post-annealing in nitrogen atmosphere produces strong low-frequency dielectric relaxation as the annealing temperature increases. While annealing in oxygen atmosphere suppresses the relaxation, but lowers the dielectric constant when the annealing temperature is high. Such results can be explained using the insulating grain boundary mechanism.

This research was sponsored by Natural Science Foundation of China (Grant No. 10204016)

**References:**

[1] C. C. Homes, T. Vogt, S. M. Shapiro, S. Wakimoto, and A. P. Ramirez, Science 2**93,** 673 (2001).

[2] M. A. Subramanian, L. Dong, N. Duan, B. A. Reisner, and A. W. Sleight, J. Solid State Chem. **151,** 323 (2000).

[3] A. P. Ramirez, M. A. Subramanian, M. Gardel, G. Blumberg, D. Li, T.Vogt, and S. M. Shapiro, Solid State Commun. **115,** 217 (2000).

[4] M. A. Subramanian and A. W. Sleight, Solid State Sci. **4,** 347 (2002).

[5] Y. J. Kim, S. Wakimoto, S. M. Shapiro P. M. Gehring and A. P. Ramirez, Solid State Commun. **121,** 625 (2002).

[6] T. B. Adams, D. C. Sinclair and A. R. West, Adv. Mater. **14**, 1321 (2002).

[7] A. Kotizsch, G. Blumberg, A. Gozar, B.Dennis, A. P. Ramirez, S. Trebst, and S. Wakimoto, Phys. Rev. B **65**, 052406 (2002)

[8] L. X. He, J. B. Neaton, M. H. Cohen, and D. Vanderbilt, Phys. Rev. B **65**, 214112 (2002)

[9] N. Kolev, R. P. Bontchev, A. J. Jacobson, V. N. Popov, V. G.Hadjiev, A. P. Litvinchuk, and M. N. Iliev, Phys. Rev. B **66**, 132102(2002)

[10] P. Lunkenheimer, V. Bobnar, A. V. Pronin, A. I. Ritus, A. A. Volkov, and A. Loidl, Phys. Rev. B **66,** 052105 (2002).

[11] D. C. Sinclair, T. B. Adams, F. D. Morrison, and A. R. West, Appl. Phys. Lett. **80**, 2153 (2002).

[12] Y. Lin, Y. B. Chen, T. Garret, S. W. Liu, C. L. Chen, L. Chen, R. P. Bontchev, A. Jacobson, J. C. Jiang, E. I. Meletis, J. Horwitz, and H. D. Wu, Appl. Phys. Lett. **81**, 631 (2002).

[13] W. Si, E. M. Cruz, P. D. Johnson, P. W. Barnes, P. Woodward, and A. P. Ramirez, Appl. Phys. Lett. **81**, 2058 (2002).

[14] M. R. Shen, S. B. Ge and W. W. Cao, J. Phys. D **34,** 2935 (2001).

[15] C. F. Yang, Jpn. J. Phys., Part 1 **36,** 188 (1997).

[16] J. B. Wu, C. W. Nan, Y. H. Lin, and Y. Deng, Phys. Rev. Lett. **89**, 217601 (2002).

[17] M. R. Shen, Z. G. Dong, Z. Q. Gan, S. B. Ge and W. W. Cao, Appl. Phys. Lett. **80**, 2538 (2002).

[18] F. M. Pontes, E. R. Leite, E. Longo, J. A. Varela, E. B. Araujo and J. A. Eiras, Appl. Phys. Lett. **76**, 2433 (2000).

**Figure Captions:**

Figure 1. X-ray diffraction patterns of the (a) CCTO target and (b) CCTO thin film.

Figure 2. The temperature dependence of (a) the dielectric constant $\varepsilon$ and (b) the loss *tanδ* for the CCTO films at different frequencies. The surface morphology of the CCTO thin film was inserted in fig.2 (a) and the cross section morphology of the CCTO thin film was inserted in fig.2 (b)

Figure 3. Frequency dependence of the dielectric constant $\varepsilon$ (a) and the loss *tanδ* (b) for (1) as-deposited Pt/CCTO/Pt capacitors, (2) post-annealed in nitrogen at 200ºC; (3) post-annealed in oxygen at 200ºC, (4) post-annealed in nitrogen at 400ºC, (5) post-annealed in oxygen at 400ºC, (6) post-annealed in nitrogen at 550ºC, and (7) post-annealed in oxygen at 550ºC.

Fig. 1 Liang Fang

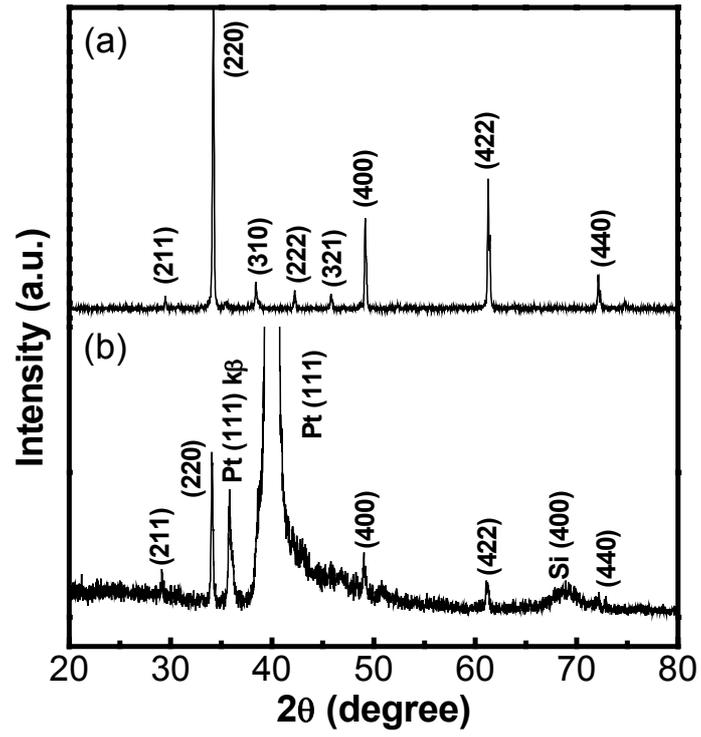

Fig. 2 Liang Fang

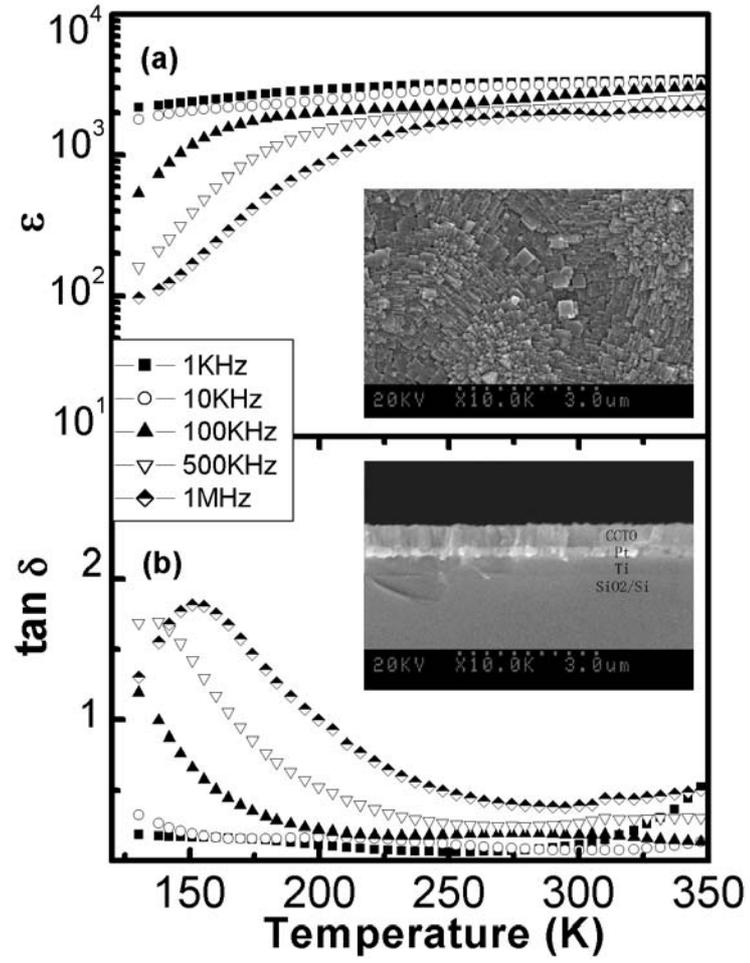

Fig. 3 Liang Fang

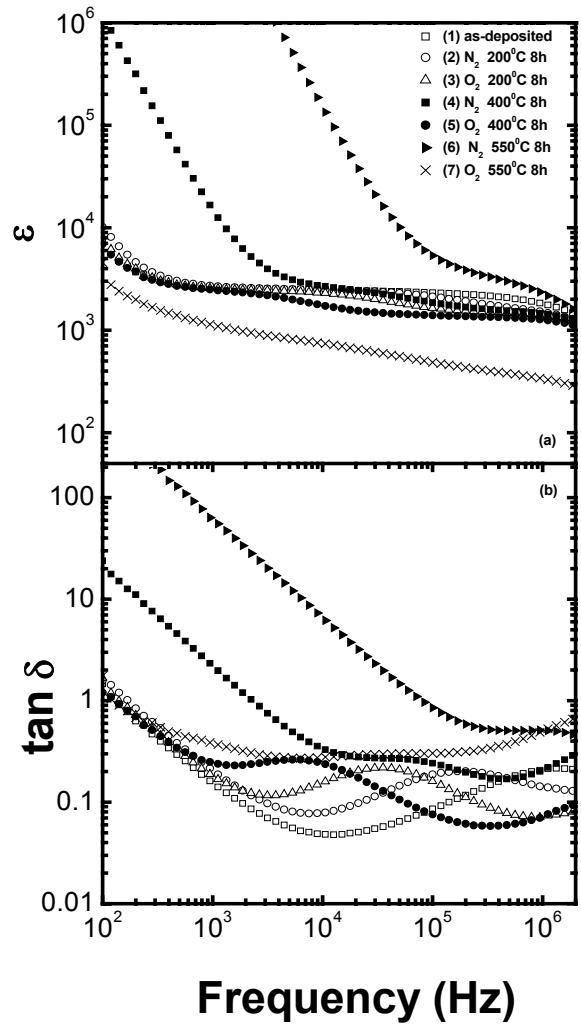